\documentclass{article}
\usepackage{graphicx}%
\def\al{&}
\def\be{\begin{equation}}
\def\ee{\end{equation}}
\def\bea{\begin{eqnarray}}
\def\eea{\end{eqnarray}}
\newcommand{\bdm}{\begin{displaymath}}
\newcommand{\edm}{\end{displaymath}}
\newcommand{\no}{\nonumber \\}

\newcommand{\ChPT}{$\chi$PT\hspace{1mm}}

\begin{document}

\vspace{3cm}
\begin{center}
{\LARGE\bf \boldmath$\pi\pi$ scattering}

\vspace{0.5cm}
H. Leutwyler\\
Institute for Theoretical Physics, University of 
Bern,\\ Sidlerstr. 5, CH-3012 Bern, Switzerland\\
$^*$E-mail: leutwyler@itp.unibe.ch
\end{center}

\begin{abstract}
Recent work in low energy pion physics is reviewed. One
  of the exciting new developments in this field is that simulations of QCD on
  a lattice now start providing information about the low energy structure of
  the continuum theory, for physical values of the quark masses. Although the
  various sources of systematic error yet need to be explored more thoroughly,
  the results obtained for the correlation function of the axial current with
  the quantum numbers of the pion already have important implications for the
  effective Lagrangian of QCD. The consequences for $\pi\pi$ scattering are
  discussed in some detail. The second part of the report briefly reviews
  recent developments in the dispersion theory of the scattering amplitude.
  One of the important results here is that the position of the lowest
  resonances of QCD can now be determined in a model independent manner and
  rather precisely. Beyond any doubt, the partial wave with $I=\ell=0$
  contains a pole on the second sheet, not far from the threshold: the lowest
  resonance of QCD carries the quantum numbers of the vacuum.
  
\vspace{0.3cm}\begin{center} Contribution to the proceedings of 
     the workshop {\it Chiral Dynamics, 

Theory \& Experiment}, Durham/Chapel Hill, NC, USA, September 2006\end{center}
\end{abstract}

\section{Introduction}
The pions are the lightest hadrons and we know why they are so light. Since
the underlying approximate symmetry also determines their basic properties at
low energy, the interaction among the pions is understood very well. In fact,
in the threshold region, the $\pi\pi$ scattering amplitude is now known to an
amazing degree of accuracy.\cite{CGL} In particular, we know how to calculate
the mass and width of the lowest resonance of QCD in a model independent
manner.\cite{CCL} The actual uncertainty in the pole position is smaller than
the estimate given in the 2006 edition of the Review of Particle
Physics,\cite{PDG 2006} by more than an order of magnitude.

The progress made in this field heavily relies on the fact that the dispersion
theory of $\pi\pi$ scattering is particularly simple: the $s$-, $t$- and
$u$-channels represent the same physical process.  As a consequence, the
scattering amplitude can be represented as a dispersion integral over the
imaginary part and the integral exclusively extends over the physical
region.\cite{Roy} Throughout the following, I work in the isospin limit
($m_u=m_d=m$ and $e=0$), where the representation involves only two
subtraction constants.\footnote{The value used for the basic QCD parameters in
  this theoretical limit is a matter of convention. It is convenient to choose
  $\Lambda_{QCD}$ and $m$ such that $F_\pi$ and $M_\pi$ agree with the
  observed values of $F_{\pi^+}$ and $M_{\pi^+}$. The result quoted by the
  PDG\cite{PDG 2006} corresponds to $F_{\pi^+}=92.42\pm 0.10\pm0.26$ MeV. The
  scattering lengths are given in units of $M_{\pi^+}$.}  These may be
identified with the $S$-wave scattering lengths $a_0^0, a_0^2$. The projection
of the amplitude on the partial waves leads to a dispersive representation for
these, the Roy equations. For a thorough discussion, I refer to
ACGL.\cite{ACGL}

The pioneering work on the physics of the Roy equations was carried out more
than 30 years ago.\cite{BFP} The main problem encountered at that time was
that the two subtraction constants occurring in these equations were not
known: if the values of $a_0^0$, $a_0^2$ are contained in the so-called
universal band -- the region spanned by the two thick lines in figure 1 below
-- the Roy equations admit a solution. Since the data available at the time
were consistent with a very broad range of $S$-wave scattering lengths, the
Roy equation ana\-ly\-sis was not conclusive.

\section{Low energy theorems for the $S$-wave scattering lengths}
The insights gained by means of chiral perturbation theory (\ChPT$\!$)
thoroughly changed the situation.\cite{Bernard Meissner 2006} The corrections
to Weinberg's low energy theorems\cite{Weinberg 1966} for $a_0^0, a_0^2$ (left
dot in figure 1) have been worked out to first non-leading order\cite{GL}
(middle dot) and those of next-to-next-to leading order are also
known\cite{BCEGS} (dot on the right).  As demonstrated in CGL,\cite{CGL} the
chiral perturbation series converges particularly rapidly near the center of
the Mandelstam triangle, so that very accurate predictions for the scattering
lengths are obtained by matching the chiral and dispersive representations
there. \begin{figure}[thb]\vspace*{-0.7cm}\hspace{-0.5cm}
\includegraphics[width=9cm,angle=-90]{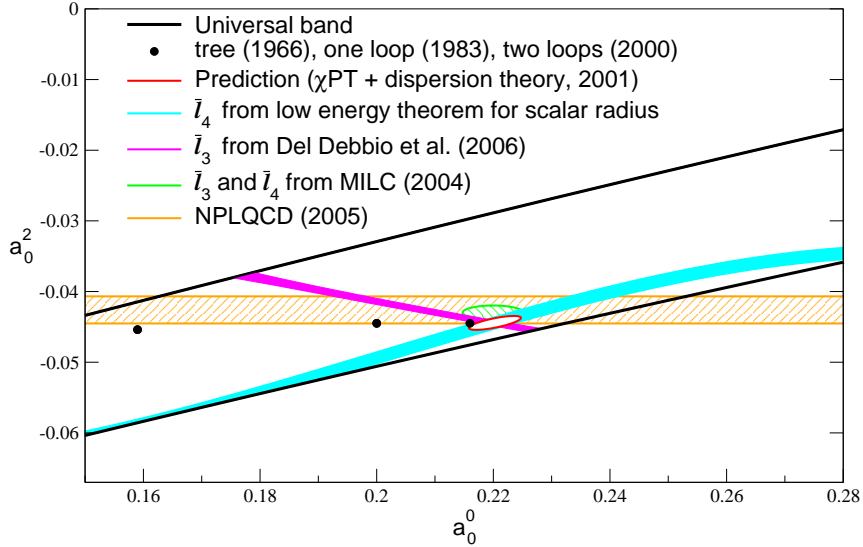}
\caption{\label{figa0a2theory}$S$ wave scattering lengths: theoretical
  results.}  
\end{figure}  
Using this method, the low energy theorems for $a_0^0$, $a_0^2$ may be brought
to the form \bea \rule{0.5cm}{0cm}2a_0^0+7a_0^2\al=\al -\,6\pi
\left(\!\frac{M_\pi}{4\pi
    F_\pi}\!\right)^{\!\!4}\!\left(\bar{\ell}_3-\frac{193}{210}\right)
+M_\pi^4 \alpha_3+O(M_\pi^6)\,,\hspace{1.2cm}(1a)\no 2 a_0^0-5 a_0^2 \al =\al
\frac{3M_\pi^2}{4 \pi F_\pi^2}+24\pi\!  \left(\!\frac{M_\pi}{4 \pi
    F_\pi}\!\right)^{\!\!4} \left(\!
  \bar{\ell}_4-\frac{887}{840}\right)+M_\pi^4\alpha_4 +O(M_\pi^6)\,.
\hspace{0.2cm}(1b)\nonumber\addtocounter{equation}{1}\eea While the chiral
expansion of $a_0^0,a_0^2$ starts at order $M_\pi^2$, with a term that is
fixed by the pion decay constant, the combination $2a_0^0+7a_0^2$ vanishes at
leading order. Two types of corrections occur at $O(M_\pi^4)$: a contribution
from the coupling constants $\bar{\ell}_3,\bar{\ell}_4$ of the effective
Lagrangian and one involving dispersion integrals, which I denote by
$M_\pi^4\alpha_3$ and $M_\pi^4\alpha_4$, respectively\footnote{In the notation
  of CGL: $\alpha_3=2\alpha_0+7\alpha_2$, $\alpha_4=2\alpha_0-5\alpha_2$.}
(these also contain higher order contributions).

A representation similar to (1) was given earlier,\cite{GL} but there, the
dispersion integrals $M_\pi^4\alpha_3,M_\pi^4\alpha_4$ were evaluated at
leading order, where they can be expressed in terms of the $D$-wave scattering
lengths. The virtue of the above form of the low energy theorems is that (a)
the neglected terms of order $M_\pi^6$ are much smaller than in the
straightforward chiral perturbation series and (b) the dispersion integrals
can be evaluated quite accurately, on the basis of the Roy equations. Since the
integrals converge very rapidly, they depend almost exclusively on the
scattering lengths $a_0^0,a_0^2$. In the vicinity of the physical values, the
linear representations \bea \rule{0.5cm}{0cm} M_\pi^4 \alpha_3 \al=\al 0.135 +
0.77\, (a^0_0-0.220) - 1.50\,(a^2_0+0.0444)\,,\hspace{1.67cm}(2a)\no
M_\pi^4\alpha_4 \al=\al 0.061 + 0.48\, (a^0_0-0.220) - 0.26\,
(a^2_0+0.0444)\,,\hspace{1.7cm}(2b)\nonumber\addtocounter{equation}{1} \eea
provide an excellent approximation. For a detailed discussion, in particular
also of the uncertainties, I refer to CGL. The estimates given there show that
the contributions of order $M_\pi^6$ in (1a) generate a correction of order
0.002, while those in equation (1b) are of order 0.003.

The low energy theorems (1) show that the $S$-wave scattering lengths are
related to the coupling constants $\bar{\ell}_3$ and $\bar{\ell}_4$. Indeed,
these play a central role in the effective theory, because they determine the
dependence of $M_\pi$ and $F_\pi$ on the quark mass at first non-leading
order: the expansion of $M_\pi^2$ and $F_\pi$ in powers of the quark mass
starts with \be M_\pi^2 =M^2\{1+\mbox{$\frac{1}{2}$}\, x\,
\bar{\ell}_3+O(x^2)\}\,, \hspace{0.5cm} F_\pi = F\{1+
x\,\bar{\ell}_4+O(x^2)\}\,,\ee where $M^2=2Bm$ is proportional to the quark
mass, $F$ is the value of the pion decay constant in the chiral limit, and
$x\equiv M^2\!/(16\pi^2 F^2)$. A crude estimate for $\bar{\ell}_3$ can be
obtained from the mass spectrum of the pseudoscalar octet:
$\bar{\ell}_3=2.9\pm 2.4$.\cite{GL} For $\bar{\ell}_4$, an analogous estimate
($\bar{\ell}_4 = 4.3\pm 0.9$) follows from the experimental value of the ratio
$F_K/F_\pi$. The low energy theorem for the radius of the scalar pion form
factor yields a more accurate result: $\bar{\ell}_4=4.4\pm0.2$.\cite{ACCGL}
The error is smaller here, because the theorem holds within
SU(2)$\times$SU(2), so that an expansion in the mass of the strange quark is
not necessary.  The small ellipse in figure \ref{figa0a2theory} shows the
prediction for the scattering lengths obtained in CGL with these values of
$\bar{\ell}_3$ and $\bar{\ell}_4$ (since the residual errors are small, the
corrections of order $M_\pi^6$ are not entirely negligible -- the curve shown
includes an estimate for these).

The narrow strip that runs near the lower edge of the universal band indicates
the region allowed if the coupling constant $\bar{\ell}_3$ is treated as a
free parameter, while $\bar{\ell}_4$ is fixed with the scalar radius, like for
the ellipse. By construction, the upper and lower edges of the strip are
tangent to the ellipse. As implied by equations (1b) and (2b), the strip
imposes an approximately linear correlation between the two scattering lengths
-- a curvature becomes visible only in the region where the theoretical
prediction would be totally wrong.

\section{Lattice results}
On the lattice, dynamical quarks can now be made sufficiently light to
establish contact with the effective low energy theory of QCD. In particular,
the MILC collaboration obtained an estimate for the coupling constants
$L_4,L_5,L_6, L_8$ of the effective chiral SU(3)$\times$SU(3)
Lagrangian.\cite{MILC} Using standard one loop formulae,\cite{GL SU3} these
results lead to $\bar{\ell}_3=0.8\pm 2.3$ and $\bar{\ell}_4=4.0\pm 0.6$. In
the plane of the two $S$-wave scattering lengths, the MILC results select the
region indicated by the second ellipse shown in figure \ref{figa0a2theory},
which is slightly larger. It would be worthwhile to analyze these data within
SU(2)$\times$SU(2), in order to extract the constants $\bar{\ell}_3$ and
$\bar{\ell}_4$ more directly. The one loop formulae of SU(3)$\times$SU(3) are
subject to inherently larger corrections from higher orders, so that the
results necessarily come with a larger error (note that, at the present state
of the art, the analysis of the lattice data relies on the one loop formulae).

The second narrow strip shown in the figure is obtained by treating
$\bar{\ell}_4$ as an unknown. Instead of showing the region that belongs to
the value of $\bar{\ell}_3$ used for the ellipse, I am taking $\bar{\ell}_3$
from a lattice study, which appeared very recently.\cite{Del Debbio et al} In
that work, the dependence of $M_\pi^2$ and $F_\pi$ on the quark mass is
investigated for two flavours of light Wilson quarks. The results for
$M_\pi^2$ are consistent with one loop \ChPT. With $M_{K\mbox{\tiny ref}}=495$
MeV, the result $\hat{\ell}_3=0.5(5)(1)$ implies $\bar{\ell}_3=3.0\pm
0.5\pm0.1$. Since this is close to the center of the estimated range, the
strip runs through the middle of the ellipse. Taken at face value, the
uncertainty in the lattice result for $\bar{\ell}_3$ is four times smaller
than the one in the estimate $\bar{\ell}_3=2.9\pm 2.4$, obtained from the
SU(3) mass formulae for the pseudoscalars, more than 20 years ago.\cite{GL}
While the width of the ellipse is dominated by the uncertainty in the value
used for $\bar{\ell}_3$, those associated with the higher order contributions
and with the phenomenological uncertainties in the dispersion integrals do
affect the width of the strip belonging to the value of Del Debbio et al.\ --
this is why the higher precision for $\bar{\ell}_3$ reduces the width by less
than a factor of four.

The coupling constants depend logarithmically on the pion mass: \be
\bar{\ell}_3= \ln \frac{\Lambda_3^2}{M_\pi^2}\,,\hspace{0.5cm} \bar{\ell}_4=
\ln\frac{\Lambda_4^2}{M_\pi^2}\,,\ee so that their value changes if the quark
mass is varied. The quoted value for $\bar{\ell}_3$ corresponds to $\Lambda_3
\simeq 560$ MeV. For $M_\pi<\Lambda_3$, the curvature term is positive and
reaches a maximum at $M_\pi=\Lambda_3/\sqrt{e}\simeq 340$ MeV. Even there,
$\frac{1}{2}x\,\bar{\ell}_3$ is less than 0.05, so that $M_\pi^2/m$ is indeed
nearly a constant. Although more work is needed to clarify all sources of
uncertainty, the calculation beautifully confirms that, to a very good
approximation, $M_\pi^2$ is linear in the quark mass. Hence the quark
condensate indeed represents the leading order parameter of the spontaneously
broken symmetry.\cite{CGL PRL} The same lattice data also yield information
about the dependence of $F_\pi$ on the quark mass, but the comparison with the
one loop formula of \ChPT is not conclusive in this case: the result for
$\bar{\ell}_4$ depends on how the data are analyzed.\cite{Del Debbio et al}

The quark mass dependence of $M_\pi$ and $F_\pi$ is also being studied by the
European Twisted Mass Collaboration. The preliminary results for the relevant
effective coupling constants read $\bar{\ell}_3= 3.5$ and $\bar{\ell}_4= 4.4$.
A thorough analysis of the uncertainties is under way.\cite{ETMC}

Finally, I mention that the scattering length of the exotic $S$-wave, $a_0^2$,
can be determined directly, from the volume dependence of the energy levels
occurring on the lattice. The horizontal band shown in figure
\ref{figa0a2theory} indicates the result obtained in this way by
NPLQCD.\cite{Beane} It is also consistent with the other pieces of information
shown in the figure. Although possible in principle, it is difficult to extend
this method to the isoscalar channel.

\section{Experiment}
Since the pions are not stable, they first need to be produced before they can
be studied, so that the experimental information is of limited accuracy.
Moreover, there are inconsistencies among the various data sets. One of the
problems is that, in most production experiments, the two pions in the final
state are accompanied by other hadrons. I know of three exceptions: production
via photons in $e^+e^-$ collisions or via $W$-bosons in the decays
$\tau^\pm\rightarrow \nu\,\pi^\pm\pi^0$ and $K\rightarrow e\nu\pi\pi$. The
first two yield excellent information about the electromagnetic and weak
vector form factors of the pion and hence also about the $P$-wave $\pi\pi$
phase shift. 
\begin{figure}[thb]\centering\vspace{-0.7cm}\hspace{-0.5cm}
\includegraphics[width=9cm,angle=-90]{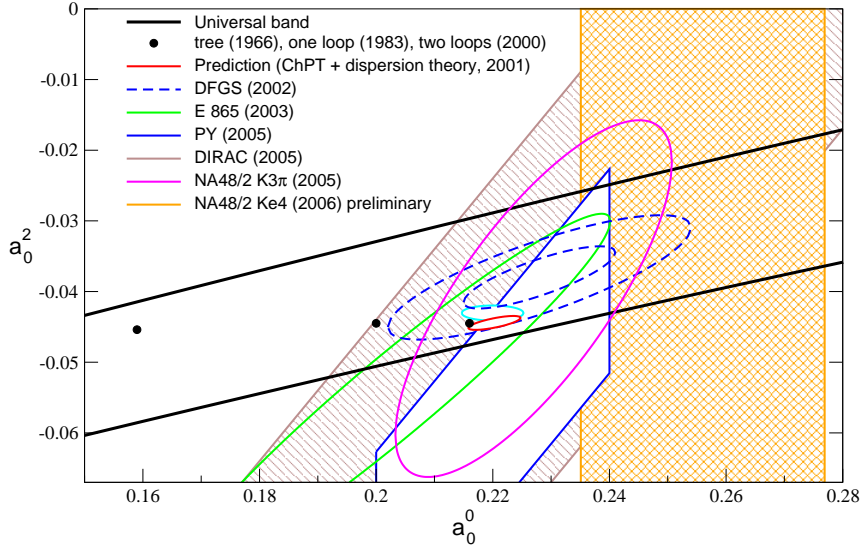}
\caption{\label{figa0a2exp}$S$ wave scattering lengths. Experimental results} 
\end{figure}
The electromagnetic form factor is of particular interest in view of the
Standard Model prediction for the magnetic moment of the muon. In this
connection, the theoretical understanding of the final state interaction among
the pions achieved in recent years can be used to reduce the experimental
uncertainties, in particular in the region below 600 MeV, where the data are
meager.\cite{Colangelo g-2} The $K$ to $\pi\pi$ transition form factors
relevant for the third process allow a measurement of the phase difference
between the $S$- and $P$-waves, $\delta_0^0-\delta_1^1$.  The ellipse labeled
E865\cite{E865} in figure 2 shows the constraint imposed on $a_0^0$ and
$a_0^2$ by the $K_{e4}$ data collected at Brookhaven. The two ellipses denoted
DFGS\cite{Stern scattering lengths} represent the 1$\hspace{0.02cm}\sigma$ and
2$\hspace{0.02cm}\sigma$ contours obtained by combining these with other
$\pi\pi$ data. The region labeled PY\cite{PY 2005} has roughly the same
experimental basis.

A very interesting proposal due to Cabibbo\cite{Cabibbo} has recently been
explored in the NA48/2 experiment at CERN: the cusp seen near threshold in the
decay $K^{\pm}\rightarrow \pi^{\pm}\pi^0\pi^0$ allows a measurement of the
difference $a_0^0-a_0^2$ of scattering lengths.\cite{NA48} In this case, the
final state involves three pions, so that the analysis of the data is not a
trivial matter.\cite{Cabibbo Isidori,CGKR,Gamiz Prades Scimemi} As seen in the
figure, the result is in good agreement with the Brookhaven data, as well as
with the theoretical predictions.  The same group has also investigated a very
large sample of $K_{e4}$ decays.\cite{NA48 ICHEP06} The result for the
scattering length $a_0^0$ is indicated by the broad vertical band. It is not
in good agreement with E865, nor with their $K\rightarrow 3\pi$ result. The
discrepancy calls for clarification.

An ideal laboratory for exploring the low energy properties of the pions
is the atom consisting of a pair of charged pions, also referred to
as pionium. The DIRAC collaboration at CERN has demonstrated that it is
possible to generate such atoms and to measure their lifetime.\cite{DIRAC}
Since the physics of the bound state is well understood,\cite{HadAtom05} there
is a sharp theoretical prediction for the lifetime.\cite{CGL} The band
labeled DIRAC in the figure shows that the observed lifetime confirms this
prediction. Pionium level splittings would offer a
clean and direct measurement of the second subtraction constant. Data on $\pi
K$ atoms would also be very valuable, as they would allow to explore the
role played by the strange quarks in the QCD vacuum.\cite{HadAtom05,Schweizer}

\section{Dispersion relations}
After the above extensive discussion of our current knowledge of the two
subtraction constants, I now wish to discuss their role in the low energy
analysis of the $\pi\pi$ scattering amplitude, avoiding technical machinery as
much as possible. Although the Roy equations represent an optimal and
comprehensive framework for that analysis, the main points can be seen in a
simpler context: forward dispersion relations.\cite{KPY} More specifically, I
consider the component of the scattering amplitude with $s$-channel isospin
$I=0$, which I denote by $T^0(s,t)$. It satisfies a twice subtracted fixed-$t$
dispersion relation in the variable $s$. In the forward direction, $t=0$, this
relation reads \bea
\label{eqFDR}
\mbox{Re}\,T^0(s,0)\al=\al c_0 + c_1\, s+ \frac{s(s-4M_\pi^2)}{\pi}\,
P\hspace{-0.3em}\int_{4M_\pi^2}^\infty\frac{dx
  \;\mbox{Im}\,T^0(x,0)} {x\,(x-4M_\pi^2)\,(x-s)}+\rule{0.5cm}{0cm}\\
\al\al\hspace{-1.7cm}+\,\frac{s(s-4M_\pi^2)}{\pi}
\int_{4M_\pi^2}^\infty\frac{dx 
  \;\{\mbox{Im}\,T^0(x,0)-3\,\mbox{Im}\,T^1(x,0)+ 5\,\mbox{Im}\,T^2(x,0)\}}
{3\,x\,(x-4M_\pi^2)\,(x+s-4M_\pi^2)}\,.\nonumber\eea The symbol $P$ indicates
that the principal value must be taken. The first integral accounts for the
discontinuity across the right hand cut, while the second represents the
analogous contribution from the left hand cut, where the components of the
scattering amplitude with $I=1,2$ also show up. According to the optical
theorem, the imaginary part of the forward scattering amplitude represents the
total cross section: in the norma\-li\-zation of ACGL,\cite{ACGL} we have
$\mbox{Im}\,T^I(s,0)=\sqrt{s(s-4M_\pi^2)}\;\sigma^I_{tot}(s)$. In this
notation, the physical total cross sections are given by
\bea
\sigma^{\pi^\pm\pi^\pm}_{tot}\al=\al \sigma^2_{tot}\,,\hspace{3.7cm}
\sigma^{\pi^\pm\pi^0}_{tot}=\mbox{$\frac{1}{2}$}
\sigma^1_{tot}+\mbox{$\frac{1}{2}$}\sigma^2_{tot}\,,\rule{1cm}{0cm}\\
\sigma^{\pi^\mp\pi^\pm}_{tot}\al=\al\mbox{$\frac{1}{3}$}\sigma^0_{tot}+
\mbox{$\frac{1}{2}$}\sigma^1_{tot}+\mbox{$\frac{1}{6}$}\sigma^2_{tot}\,,
\hspace{1.1cm}  
\sigma^{\pi^0\pi^0}_{tot}=\mbox{$\frac{1}{3}$}
\sigma^0_{tot}+\mbox{$\frac{2}{3}$}\sigma^2_{tot}\,.\nonumber\eea 
As mentioned already, the subtraction term is determined by $a_0^0,a_0^2$:  \be
c_0+c_1\, s = 32\,\pi
\left\{a_0^0+(2a_0^0-5a_0^2)\,\frac{s-4M_\pi^2}{12M_\pi^2}\right\}\,.\ee

A dispersion relation of the above type also holds for other processes. What
is special about $\pi\pi$ is that the contribution from the crossed channels
can be expressed in terms of observable quantities -- total cross sections in
the case of forward scattering. The contribution from the left hand cut is
dominated by the $\rho$-meson, which generates a pronounced peak in the total
cross section with $I=1$. This contribution is known very accurately from the
process $e^+e^-\rightarrow\pi^+\pi^-$. In the physical region, $s> 4M_\pi^2$,
the entire contribution from the crossed channels is a smooth function that
varies only slowly with the energy. Note, however, that this contribution is
by no means small.\cite{Zhou} 

The angular momentum barrier suppresses the higher partial waves: at low
energies, the first term in the partial wave decomposition \bea\label{eqPWE}
\mbox{Re}\,T^0(s,t)\al=\al 32 \pi \left\{\mbox{Re}\,t^0_0(s)+5\,P_2(z)\,
  \mbox{Re}\,t^0_2(s)+\ldots\right\}\\
t\al=\al \mbox{$\frac{1}{2}$}(4M_\pi^2-s)(1-z)\nonumber\eea represents the
most important contribution. In the vicinity of the threshold, where the
contribution from the higher angular momenta is negligibly small, the
dispersion relation (\ref{eqFDR}) thus amounts to an expression for the real
part of the isoscalar $S$-wave. For brevity, I refer to this partial wave as
$S^0$.  

Equation (\ref{eqFDR}) imposes a very strong constraint on $S^0$, for the
following reason. Suppose that all partial waves except this one are known.
The relation then determines Re$\,t^0_0(s)$ as an integral over
Im$\,t^0_0(x)$.  In the elastic region, unitarity already fixes the real part
in terms of the imaginary part -- we thus have two equations for the two
unknowns Re$\,t^0_0(s)$ and Im$\,t^0_0(s)$.  Accordingly, we may expect that,
in the elastic region, the f.d.r.\ unambiguously fixes the $S^0$-wave. Indeed,
this is borne out by the calculation, which is described in some detail
elsewhere.\cite{Azores}

One of the main results established in CGL is that the Roy equations fix the
behaviour of the $S^0$-wave below 800 MeV almost entirely in terms of
three parameters: the two subtraction constants $a_0^0, a_0^2$ and the value
of the phase at 800 MeV. In particular, the behaviour of the scattering
amplitude at high energies is not important. This can also be seen on the
basis of equation (\ref{eqFDR}): replacing our representation of the
scattering amplitude above $K\bar{K}$ threshold by the one proposed in
KPY,\cite{KPY} for instance, the solution of the f.d.r.\ very closely follows
the solution of the Roy equations that belongs to the same value of the phase
at 800 MeV.\cite{Azores}

\section{The lowest resonances of QCD}
The positions of the poles in the S-matrix represent universal properties of
the strong interaction, which are unambiguous even if the width of the
corresponding resonance turns out to be large,\cite{Universality of poles} but
they concern the non-perturbative domain, where an analysis in terms of the
local degrees of freedom of QCD -- quarks and gluons -- is not in sight. First
quenched lattice explorations of the pole from the $\sigma$ have
appeared,\cite{Liu} but in view of the strong final state interaction in this
channel, it will take some time before this state can reliably be reached on
the lattice.

One of the reasons why the values for the pole position of the $\sigma$ quoted
by the Particle Data Group cover a very broad range is that all of these rely
on the extrapolation of hand made parametrizations: the data are represented
in terms of suitable functions on the real axis and the position of the pole
is determined by continuing this representation into the complex plane. If the
width of the resonance is small, the ambiguities inherent in the choice of the
parametrization do not significantly affect the result, but the width of the
$\sigma$ is not small.

A popular approach to the problem is based on the so-called {\it inverse
  amplitude method}.  Applying it to the \ChPT representation of the
scattering amplitude\cite{IAM} invariably produces a pole in the right ball
park. The procedure definitely improves the quality of the two-loop
approximation of \ChPT on the real axis, because it respects unitarity in the
elastic region. The extrapolation into the complex plane, however, also
contains a number of fake singularities. Like all other parametrizations, this
approach relies on a model. I do not know of a way to estimate the systematic
uncertainties generated if QCD is replaced by one model or the other -- if
error estimates are given at all, these necessarily rely on guesswork. A
thorough discussion of the problems inherent in the IAM approach appeared
several years ago.\cite{Zheng Pade}

We have found a method that does not require a parametrization of the data at
all. It relies on the fact that (a) the $S$-matrix has a pole on the second
sheet if and only if it has a zero on the first sheet, (b) the Roy equations
are valid not only on the real axis, but in a limited domain of the first
sheet, (c) the poles from the lowest resonances, $\sigma$, $\rho$, $f_0(980)$,
all occur in that domain. The numerical evaluation of the pole position is
straightforward.  The one closest to the origin occurs at\cite{CCL}
\be\label{eqmsigma} M_\sigma-\frac{i}{2}\,\Gamma_\sigma =\sqrt{s_\sigma}=441\,
\rule[-0.2em]{0em}{1em}^{+16}_{-\,8}-
\,i\;272\,\rule[-0.2em]{0em}{1em}^{+\,9}_{-12.5}\;\mbox{MeV}\,,\ee where the
error accounts for all sources of uncertainty. We may, for instance, replace
our representation for $S^0$ by the one proposed by Bugg\cite{Bugg sigma pole}
or replace the entire scattering amplitude by the parametrization in
KPY.\cite{KPY} In either case, the outcome for the pole position is in the
above range. For more details, I refer to CCL\cite{CCL} and to a recent
conference report.\cite{Krakow}

In the meantime, the method described above was applied to the case of $\pi K$
scattering, with the result that the lowest resonance in that channel occurs
at $m_\kappa =(658\pm 13)-i\,(278.5\pm 12)$ MeV.\cite{Descotes-Genon and
  Moussallam 2006} Evidently, the physics of the $\kappa$ is very similar to
the one of the $\sigma$.

\section{On the working bench} 
We are currently extending the work described in CGL to higher energies. The
price to pay is that the contributions from the high energy region then become
more important.  Since the first few terms of the partial wave expansion do
not represent a decent approximation there, one instead uses a Regge
representation for the asymptotic domain. In CGL, we borrowed that from
Pennington.\cite{Pennington Annals} In the meantime, we have performed a new
Regge analysis, invoking experimental information as well as sum rules to pin
down the residue functions. 
\begin{figure}[thb]\centering\vspace{-0.3cm}
\includegraphics[width=4.4cm,angle=-90]{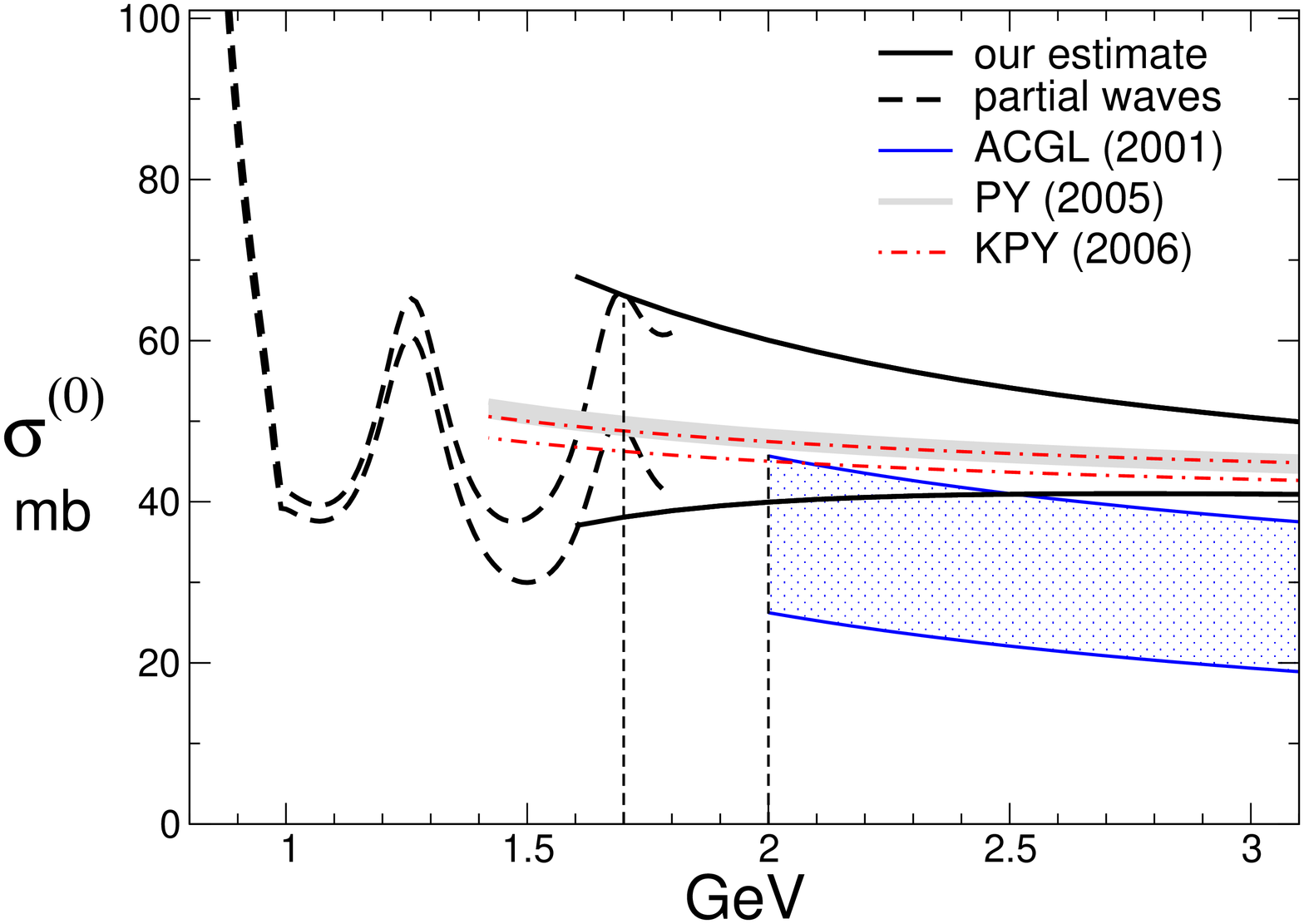}\hspace{-0.2cm}
\includegraphics[width=4.4cm,angle=-90]{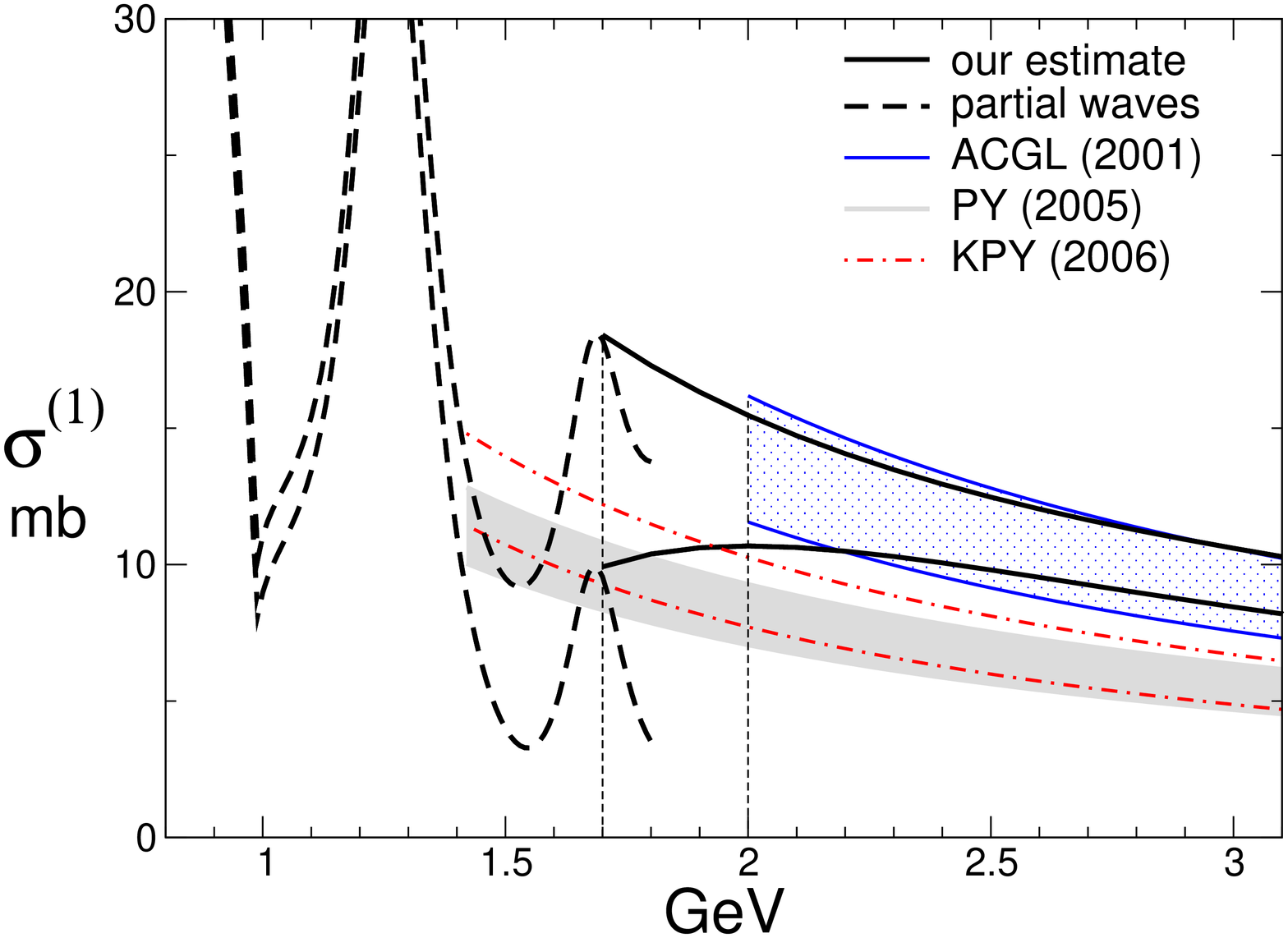}
\caption{\label{figRegge}Asymptotic behaviour of the total cross sections with
  $I_t=0$ and $I_t=1$.}  
\end{figure}  
Brief accounts of this work were given
elsewhere.\cite{Irinel Regge} Figure \ref{figRegge} shows the preliminary
results obtained for the total cross sections with $t$-channel isospin 0 and
1: \be \sigma_{tot}^{(0)}=\mbox{$\frac{1}{3}$}\sigma_{tot}^0+\sigma_{tot}^1+
\mbox{$\frac{5}{3}$}\sigma^2_{tot}\,,\hspace{0.2cm}\sigma_{tot}^{(1)}=
\mbox{$\frac{1}{3}$}\sigma_{tot}^0+
\mbox{$\frac{1}{2}$}\sigma_{tot}^1-\mbox{$\frac{5}{6}$}\sigma^2_{tot}\,.\ee We
include pre-asymptotic contributions and assume that the Regge parametrization
yields a decent approximation above 1.7 GeV.  At lower energies, the
uncertainties in that representation become larger than those in the sum over
the partial waves, which is also shown. The comparison with the representation
used in ACGL confirms the cross section with $I_t=1$, while the one for
$I_t=0$ comes out larger by 1 or $2\hspace{0.05cm}\sigma$.  In the solutions
of the Roy equations, the effects generated by such a shift are barely visible
below 800 MeV. We hope to complete the analysis soon, as well as the
application to the electromagnetic form factor of the pion, for which an
accurate representation is needed in connection with the Standard Model
prediction for the magnetic moment of the muon.

It is a pleasure to thank the organizers of the meeting for their kind
hospitality during a very pleasant stay at Chapel Hill and Claude Bernard,
Leonardo Giusti and Steve Sharpe for useful comments about the text. Also, I
acknowledge Balasubramanian Ananthanarayan, Irinel Caprini, 
Gilberto Colangelo and J\"urg Gasser for a most enjoyable and fruitful
collaboration -- the present report is based on our common work.

\end{document}